\pdfoutput=1

\documentclass[11pt]{article}

\usepackage{acl}

\usepackage{times}
\usepackage{latexsym}
\usepackage{amsfonts}
\usepackage{CJK}

\usepackage[T1]{fontenc}

\usepackage[utf8]{inputenc}
\usepackage{amsmath} 
\usepackage{microtype}

\usepackage{inconsolata}

\usepackage{graphicx}%
\usepackage{tabulary}
\usepackage{multirow}%
\usepackage{amsmath,amssymb,amsfonts}%
\usepackage{amsthm}%
\usepackage{mathrsfs}%
\usepackage[title]{appendix}%
\usepackage{xcolor}%
\usepackage{textcomp}%
\usepackage{manyfoot}%
\usepackage{booktabs}%
\usepackage{algorithm}%
\usepackage{algorithmicx}%
\usepackage{algpseudocode}%
\usepackage{listings}%
\usepackage{colortbl}
\usepackage{diagbox}
\usepackage{multirow}
\usepackage{times}
\usepackage{multicol}
\usepackage{arydshln} 
\usepackage{booktabs}
\usepackage{tabularx}
\usepackage{float}
\usepackage{microtype}
\usepackage{inconsolata}
\usepackage{verbatim}
\usepackage{array}
\usepackage[caption=false,font=footnotesize,labelfont=rm,textfont=rm]{subfig}
\usepackage{textcomp}
\usepackage{stfloats}
\usepackage{url}
\usepackage[T1]{fontenc}
\usepackage{makecell}
\usepackage{ragged2e}
\usepackage{listings}

\newcommand{\ie}{\textit{i.e.}}

\title{Towards Real-World Stickers Use: A New Dataset for Multi-Tag \\Sticker Recognition}

\author{
Bingbing Wang, Bin Liang, Chun-Mei Feng$^{*}$,  Wangmeng Zuo, Zhixin Bai, \\ \textbf{Shijue Huang},  \textbf{Kam-Fai Wong}, \textbf{Xi Zeng}, \textbf{Ruifeng Xu}$^{*}$\\
}

\begin{document}
\maketitle
\begin{abstract}
In real-world conversations, the diversity and ambiguity of stickers often lead to varied interpretations based on the context, necessitating the requirement for comprehensively understanding stickers and supporting multi-tagging.
To address this challenge, we introduce \textbf{StickerTAG}, the first multi-tag sticker dataset comprising a collected tag set with 461 tags and 13,571 sticker-tag pairs\footnote{We believe that the release of this dataset will provide exciting research opportunities and encourage further research in sticker analysis. }, designed to provide a deeper understanding of stickers. 
Recognizing multiple tags for stickers becomes particularly challenging due to sticker tags usually are fine-grained attribute aware. 
Hence, we propose an Attentive Attribute-oriented Prompt Learning method, \ie, \textbf{Att$^2$PL}, to capture informative features of stickers in a fine-grained manner to better differentiate tags. 
Specifically, we first apply an Attribute-oriented Description Generation (ADG) module to obtain the description for stickers from four attributes.
Then, a Local Re-attention (LoR) module is designed to perceive the importance of local information. 
Finally, we use prompt learning to guide the recognition process and adopt confidence penalty optimization to penalize the confident output distribution.
Extensive experiments show that our method achieves encouraging results for all commonly used metrics.

\end{abstract}

\section{Introduction}
The rapid development of stickers has rendered them indispensable tools in our daily lives, seamlessly integrating into various aspects of our routine communications and digital expressions. Numerous research endeavors have been dedicated to sticker-based multi-modal sentiment analysis \cite{14-ge2022towards,16-zhao2023sticker820k}. However, compared to abstract sentiment labels such as positive, negative, and neutral, the detailed information provided by sticker tags helps users quickly select stickers that match the current conversation, thereby enhancing the user experience.

\begin{figure}[!t]
  \centering
  \includegraphics[width=\linewidth]{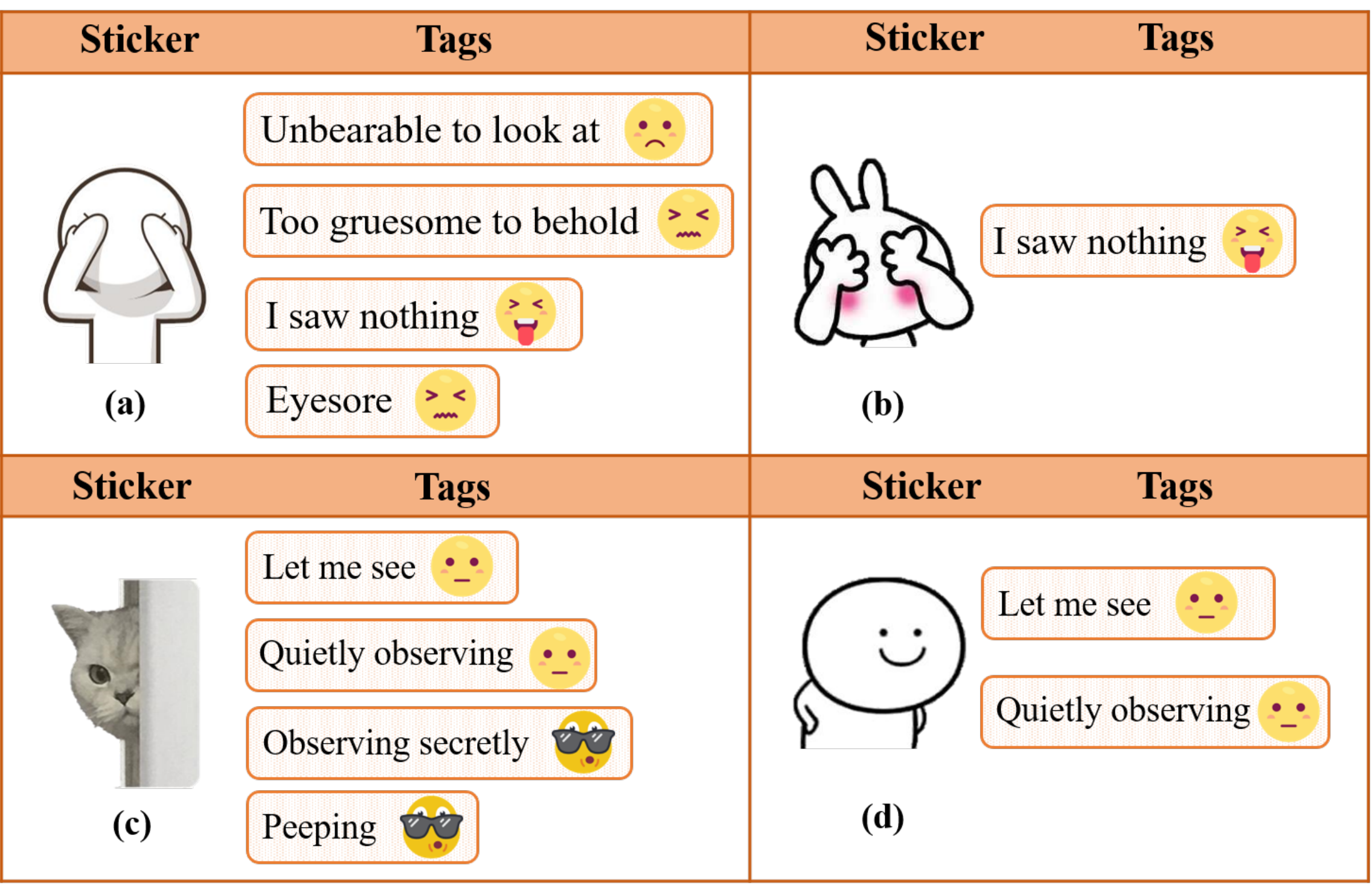}
  \caption{Examples of stickers along with multiple tags.  
  }
  \label{f1}
  \vspace{-0.3cm}
\end{figure}


While most messaging apps, such as WeChat, Line, Telegram, and WhatsApp, assign only a tag to each sticker, the same sticker may be interpreted differently by users in real-world conversations.
For instance,  Figure \ref{f1} (a) shows that a cartoon role covering its eyes can convey more than one meaning with multiple emotions, including \{\textit{Unbearable to look at; Too gruesome to behold; I saw nothing; Eyesore}\}.
Therefore, we introduce a multi-tag sticker dataset called \textbf{StickerTAG}. 
It is the first tag-annotated sticker dataset, comprising $13,571$ sticker-tag pairs and a collection of $461$ tags. 
A thorough and stringent process is devised for tag construction and sticker annotation.
We will release our dataset to encourage research on stickers in more practical scenarios.

Multi-tag sticker recognition is inherently challenging due to the need to precisely discern the rich and subtle meanings encapsulated within stickers. 
For instance, both characters in Figure \ref{f1} (a) and (b) are depicted covering their faces, but in Figure \ref{f1} (b), the cartoon rabbit is blushing while doing so, which can be best described by the tag "\textit{I saw nothing}".
Furthermore, while both "\textit{Observing secretly}" and "\textit{Peeping}" imply the act of looking, their appropriate application is limited to the specific action in Figure \ref{f1} (c), where a cat is stealthily peeking from behind a wall, rather than (d). 
This scenario highlights the importance of identifying subtle cues within the sticker's imagery to distinguish between closely related concepts accurately.

To tackle this challenge, we design an Attentive Attribute-oriented Prompt Learning method, abbreviated as \textbf{Att$^2$PL}, which begins with an Attribute-oriented Description Generation (ADG) module powered by a Multi-modal Large Language Model (MLLM) to derive the description of stickers based on the content, style, role, and action.
Furthermore, 
the sticker is fed into a Local Re-attention (LoR) module 
for 
deriving patch-attentive embedding.
Then, the recognition process is guided by prompt classification, where we initialize the soft prompt with attribute-oriented descriptions. 
Finally, we implement a confidence penalty optimization for precise and reliable recognition.
We employ commonly used metrics to evaluate the performance of our model, and extensive experiments conducted on our StickerTAG dataset show that our approach significantly outperforms the strong baselines.
Our contributions are summarized as follows:
\begin{itemize}
    \item To our knowledge,  StickerTAG is the first multi-tag sticker dataset.  It includes 461 tags and 13,571 sticker-tag annotations towards more practical scene. 
    \item We propose an Attentive Attribute-oriented Prompt Learning approach, facilitating the capture of fine-grained features of stickers.
    \item  Experiments conducted on the StickerTAG dataset show that our model outperforms all baselines, clearly confirming its effectiveness.
\end{itemize}

\section{Related Work}
\subsection{Sticker Dataset} 
Stickers are predominantly utilized to enhance emotional and expressive communication in chat conversations and social media, thereby fostering greater engagement. 
CSMSA \citep{14-ge2022towards} is the pioneering work in multi-modal sentiment analysis focusing on stickers. To obtain multi-modal data comprising both text and stickers, the approach iterates through each sticker in the chat history, collecting the corresponding context.
Considering the lack of sticker emotion data, \citet{15-liu2022ser30k} collected a sticker emotion recognition dataset named SER30K. It consists of a total of 1,887 sticker themes with a total of 30,739 sticker images. 
\citet{16-zhao2023sticker820k} introduced a sizable Chinese dataset, Sticker820K, comprising 820k image-text pairs. 
Nevertheless, stickers may be interpreted differently by users. The above datasets assign a single sentiment label to each sticker, limiting the ability to comprehensively capture the diverse information that a sticker might convey. 

\subsection{Sticker-based Method} 
Unlike image-text pairs in vision-language datasets, stickers demand finer-grained emotion recognition and subject understanding due to their increased diversity and domain specificity. 
This has prompted the development of specialized approaches tailored to the unique attributes of stickers, as relying on generic vision-language models (VLMs) may prove suboptimal.
CSMSA \citep{14-ge2022towards} introduces a Sticker-Aware Multi-modal Sentiment Analysis Model to tackle challenges in multimodality, inter-series variations, and multi-modal sentiment fusion. 
\citet{16-zhao2023sticker820k} delved into aligning visual and textual features with emotional cues and artificial painting. It investigates the feasibility of seamlessly integrating additional tools into Large Language Models (LLMs). 
However, simple classification methods are insufficient to address the challenges posed by multi-tag sticker recognition. This prompts us to delve into more fine-grained features of stickers and enhances the ability of the model to differentiate between tags more effectively. 


\section{StickerTAG Dataset}

In this section, we provide a detailed description of our dataset construction process. Firstly, we discuss the process of tag construction, illustrated in Section \ref{Tag Construction}. Next, we outline the procedure for collecting and annotating stickers in Section \ref{Sticker Collection and Annotation}. Finally, we analyze the characteristics of our dataset in Section \ref{Characteristics}.


\subsection{Tag Construction} 
\label{Tag Construction}
We start by compiling all keywords from a well-known Chinese sticker repository\footnote{https://github.com/zhaoolee/ChineseBQB}. Next, we employed the K-means clustering algorithm \citep{kmeans} for cluster analysis. To prepare the text for analysis, we segmented it into words using Jieba, removed stop words and special characters, and then extracted features by converting the text into numerical feature vectors using the term frequency-inverse document frequency (TF-IDF).
Ultimately, we determined an appropriate number of clusters using the Elbow Method \citep{elbow}. Given the extensive list of keywords, we initially defined a range with a step size of 100. Through the elbow method, a notable turning point is identified at a K value of 400, prompting us to narrow down the range to [400, 500]. Subsequently, with a refined step size of 1, we repeated the elbow method to determine the optimal K value. 
 
To ensure the credibility and accuracy of tags, a team of experts with relevant knowledge is engaged to assign uniform names to each cluster. They carefully reviewed and revised the final results to create a comprehensive tag set, confirming their applicability. Revision or addition of new tags may be necessary during this process to ensure the completeness of the tag set.

\begin{figure*}[!t]
  \centering
  \includegraphics[width=\linewidth]{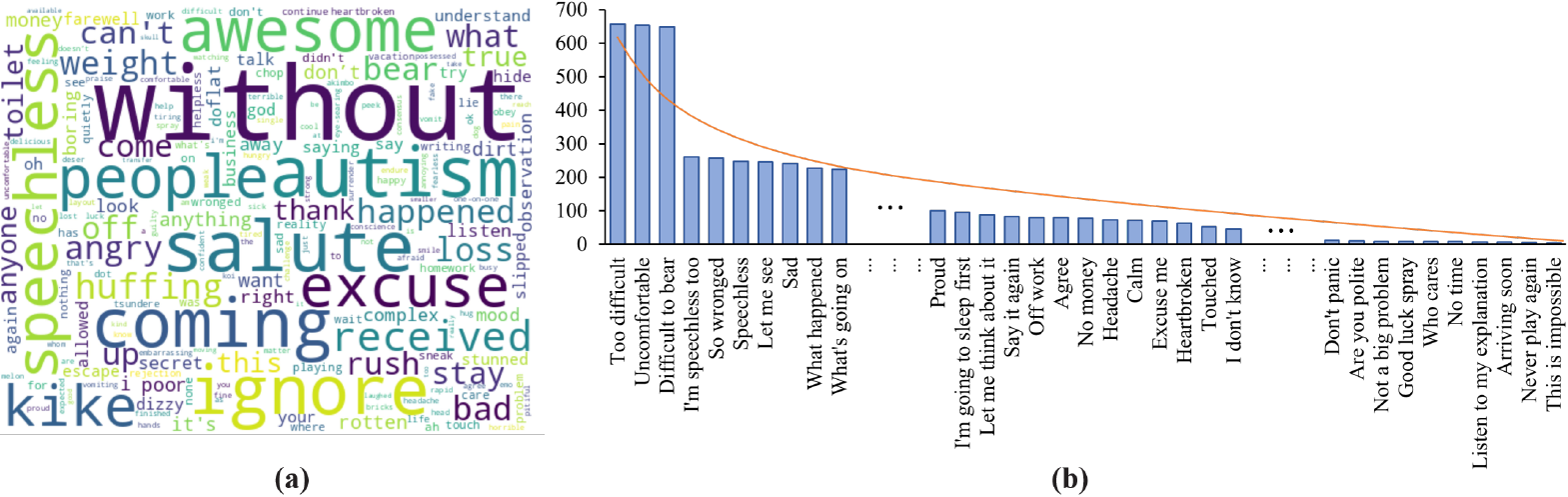}
  \caption{(a) Word cloud distribution of the sticker tags. Larger text size indicates a higher frequency of occurrence. (b) Number of samples per tag, highlighted by an orange trend line.}
  \label{f2}
  \vspace{-0.2cm}
\end{figure*}

\subsection{Sticker Collection and Annotation}
\label{Sticker Collection and Annotation}
\textbf{Sticker Collection.}
After obtaining the tag set, we collected stickers by searching for keywords using input methods such as Sogou 
and Baidu. 
We implemented strict rules for filtering candidate images, extracting key frames from dynamic formats like GIF and APNG, and saving them as JPG. Subsequently, we enlisted numerous experts to filter out stickers containing inappropriate, offensive, or insulting content. The resulting dataset comprises static image formats such as PNG and JPG.


\textbf{Annotation.} 
A sticker image encapsulates multiple cues for expression, and the tags are closely tied to the subjective perceptions of annotators who may assign varied tags to the same sticker based on their social backgrounds. 
To address this variability, we invited three annotators to label the tags for each sticker, assigning each sticker to all annotators. A tag is considered effective only if at least two annotators provide the same annotation. 
Stickers for which all three annotators provide inconsistent tags are extracted separately for further discussion.

\subsection{Characteristics} 
\label{Characteristics} 

The final dataset contains 461 tags and 13,571 sticker-tag pairs annotated data, and
the large-scale dataset is for future work.
This dataset is the first annotated sticker dataset for multi-tag recognition to the best of our knowledge.
We present the word cloud distribution of sticker tags in Figure \ref{f2} (a). We can conclude that certain tags prominently feature emotionally charged and conversational words, reflecting a diverse range of topics.
Moreover, Figure \ref{f2} (b) shows the number samples per tag. It can be seen that our dataset exhibits a long-tailed distribution like most of the multi-label datasets.
We also conduct studies on the tags of each sticker. According to our statistics, nearly 0.35\% stickers have 6 tags, 2.67\% of 5 tags, 9.23\%  of 4 tags,  14.99 \% of 3 tags, 25.31\%  of 2 tags, and  47.45\%  of 1 tag. 
And the average tag length is 3.10 words.

\begin{figure*}[!t]
  \centering
  \includegraphics[width=0.9\linewidth]{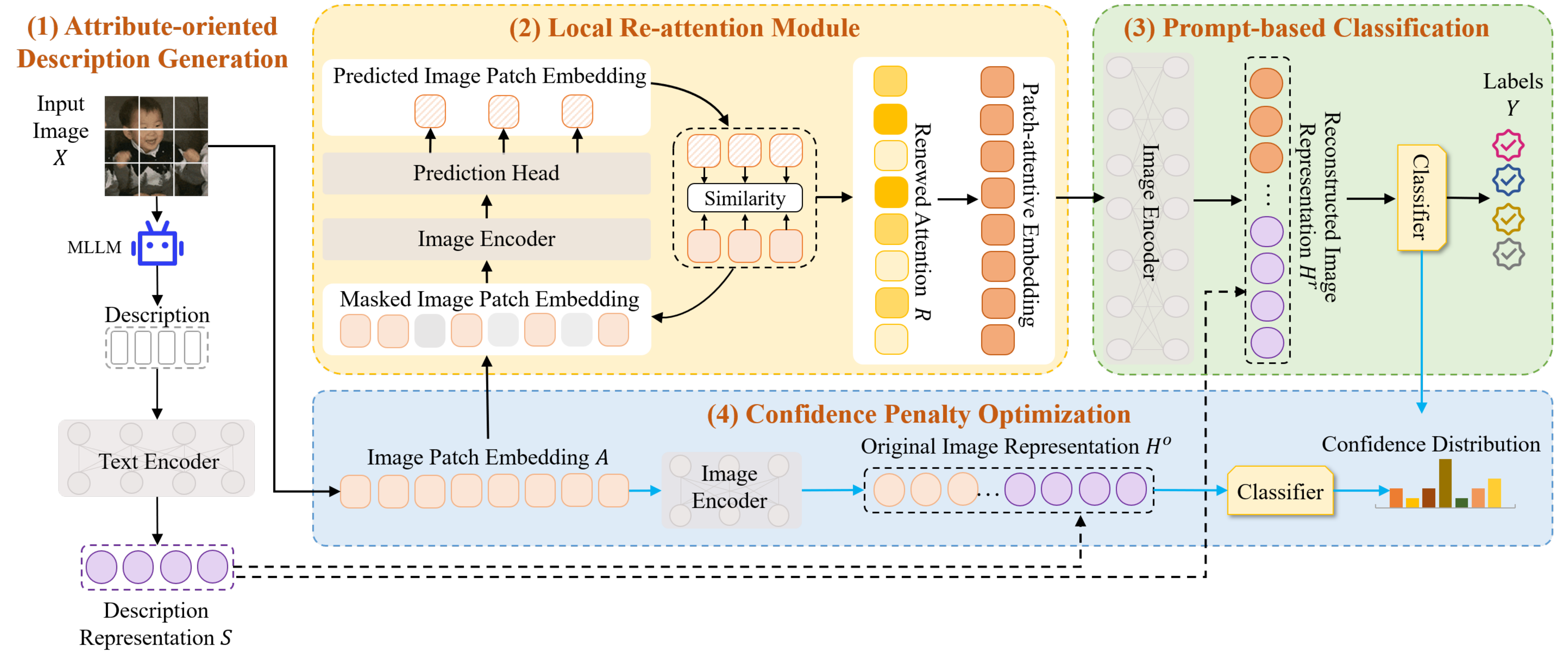}
  \caption{Illustration of the proposed Att$^2$PL method comprising (1) Attribute-oriented Description Generation, (2) Local Re-attention Module, (3) Prompt-based Classification, and (4) Confidence Penalty Optimization (blue lines).}
  \label{f3}
  \vspace{-0.2cm}
\end{figure*}

\section{Method}
This section begins with a concise introduction to the multi-tag sticker recognition task, 
followed by a detailed explanation of our proposed Att$^2$PL framework.
As demonstrated in Figure \ref{f3}, Att$^2$PL consists primarily of four components:
1) \textbf{Attribute-oriented Description Generation} designs prompts based on the attributes of the stickers for MLLM to generate attribute-oriented descriptions. 
2) \textbf{Local Re-attention Module} utilizes masked images in a novel manner to derive renewed attention for deriving patching-attentive embedding. 
3) \textbf{Prompt-based Classification} exploits attribute-oriented descriptions to initialize the soft prompts and combines patch-attentive embedding to achieve the multi-tag prediction. 
4) \textbf{Confidence Penalty Optimization} penalizes confident output distributions for implicit feedback of renewed attention.

\subsection{Task Definition}
Multi-tag sticker recognition is a task to predict a set of tags for an input sticker. Formally, given a set of stickers $\mathcal{X}= \{X_1,X_2,...,X_n\}$, and $\mathcal{Y}=\{Y_1,Y_2,...,Y_m\}$ be a ground truth set of tags, a multi-tag pattern can be defined as a pair $(x,y)$, where $n$ and $m$ are the number of stickers and tags. $x \in \mathcal{X}$ indicates a sticker and $y \subseteq \mathcal{Y}$ demonstrates the tag set. The goal of the multi-tag sticker recognition task is to construct a classifier $f$ to predict a set of tags given a sticker: $\hat y = f(x)$.

\subsection{Attribute-oriented Description Generation}
Recently, as the era of large language models unfolds, MLLMs are also emerging, bringing new opportunities for analyzing and generating textual and visual information \citep{llava,minigpt4,44-bai2023qwen}. Nevertheless, most MLLM severely suffer from object hallucination and are even more prone to hallucinating than small vision-language. In this way, as shown in Figure \ref{f4}, we elaborately design a prompt that is based on capturing information more precisely in four aspects: content, style, role, and action. This aims to reduce unnecessary interference from irrelevant information, especially when dealing with stickers.

We use a multi-turn interaction approach to prompt MLLM-generating utterances with specific intents. We first use several turns of interactions, including the system prompt like “This is a sticker used in conversation...” to simulate the utterance generation ability of MLLM. Then to further describe the details of this task, we guide the MLLM to generate specific information step by step:

\begin{figure}[!t]
  \centering
  \includegraphics[width=0.9\linewidth]{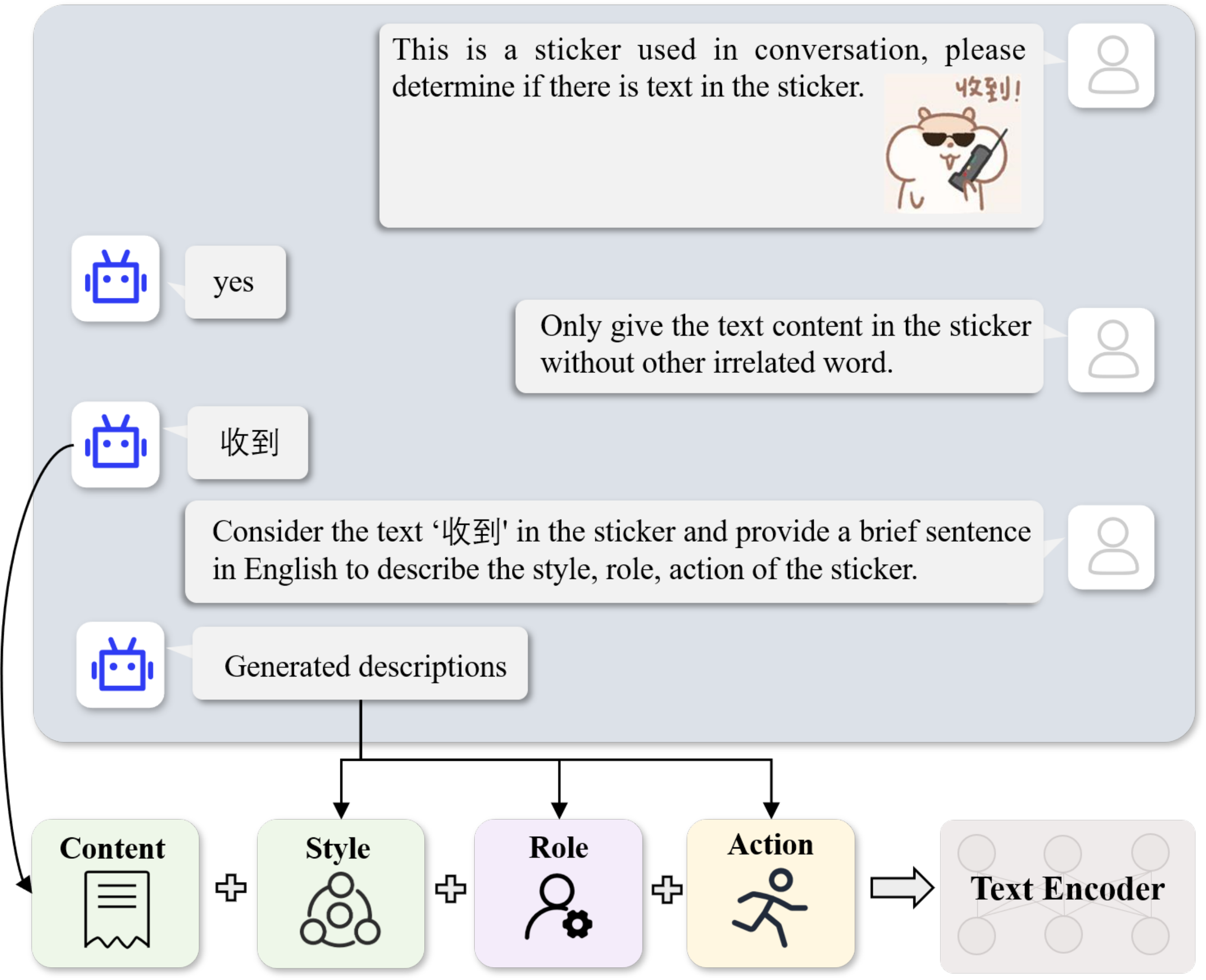}
  \caption{Overview of attribute-oriented description generation. }
  \label{f4}
  \vspace{-0.2cm}
\end{figure}


\begin{itemize}

\setlength{\itemsep}{2pt}
\setlength{\parsep}{2pt}
\setlength{\parskip}{2pt}
    \item[-] Please determine if there is text in the sticker. 
    \item[-] Only give the text content in the sticker without other unrelated words.
    \item[-] Consider the text in the sticker and provide a brief sentence in English to describe the style, role, and action of the sticker.
\end{itemize}

Then, a text encoder is employed to acquire the description representations for the description of content, style, role, and action.

\subsection{Local Re-attention Module}
The challenge for multi-tag recognition is to guide the model to focus on crucial region information within stickers.
Intuitively, we introduce a local re-attention module (LoR) implemented through masked image modeling to capture significant region information.

\textbf{Patch-aligned random masking.} Given an input sticker $X \in \mathbb{R}^{C \times H \times W}$, we first divide the sticker image into patches $X=\{x_1,x_2,...,x_N\}, x_i \in \mathbb{R}^{P^2C}$, serving as fundamental processing units for vision transformation. Where $C$, $H$, and $W$ are the numbers of channels, height, and width of $X$, $N=HW/P^2$ represents the number of patches, $P$ denotes the patch size. 
Subsequently, we tokenize patches into $N$ visual tokens to derive patch embedding $A=\{a_1,a_2,...,a_N\}$.
We randomly mask approximately $L$ patches, with the masked positions denoted as $M \subseteq \{1,...,N\}^L$. 
Then we employ a learnable mask token embedding $e_{[M]} \in \mathbb{R}^D$ to replace each masked patch, where $D=P^2C$ represents the dimension of the mask token vector. 
The corrupted image patches $x^M=\{x_i:i \notin M\}^N_{i=1} \cup \{e_{[M]} \in \mathbb{R}^D\}^N_{i=1}$ are then input into 
an image encoder to obtain final hidden vectors.

\textbf{Prediction Head.} The prediction head with a linear layer will be applied to the final hidden vectors to produce the prediction of the masked area. 
We initially predict raw pixels for the masked area through regression for each masked position. Subsequently, we map each feature vector back into a feature map at the original resolution. This vector is then responsible for predicting the corresponding raw pixels, resulting in the predicted image patch embedding ${\hat x_i}^L, i \in M$.
Furthermore, we compute the similarity between each predicted image patch embedding and the corresponding original patch embedding as ${a_i}^L, i \in M$ for each patch. 
After multiple iterations, we can obtain a feature vector consisting of similarity scores for each patch, denoted as $\{r_i\}_1^N$, and then transform it into a renewed attention represented as $R=\{\hat r_1, \hat r_2,..., \hat r_N\}$, where $\hat r_i = 1-r_i$. 
Subsequently, we multiply $R$ with the patch embedding $A$ to derive patch-attentive embedding.

\subsection{Prompt-based Classification}
We pass the patch-attentive embedding through the image encoder to get the patch-attention representation $h$, then append learnable soft prompts to obtain the reconstructed image representation $H^{r}$, which is designed as: 
\begin{equation}
    H^{r} = \text{[CLS]} S_1 \cdots S_k h \text{[SEP]}.
\end{equation}
where $S_k, k\in\{1,..., 4\}$ denotes learnable vectors with the same dimension as the word embedding, which are initialized by four description representations. 
To perform multi-tag recognition, we employ softmax as a classifier to make the final prediction. $H^{r}$ is input to the softmax layer, and the probability of each tag is estimated. 
The objective function can be described as follows:
\begin{equation}\small
    \mathcal{L}_{main}=-\frac{1}{n}\sum_{i=1}^n \sum_{j=1}^m 1\{y_i=j\} \ln \frac{e_r^{\theta_j^\top H_i^{r}}}{\sum_{i=1}^m e_r^{\theta_i^\top H_i^{r}}} .
\end{equation}
where the indicator function equals $1\{\cdot\}$ when $H_i^{r}$ holds tag $j$ and equals 0 otherwise, $p_j^r = e_r^{\theta_j^\top H^{r}_i}$ is the reconstructed confidence distribution.
The model parameter $\theta$ can be obtained by minimizing $\mathcal{L}_{main}$. 
Finally, the top-C predicted tag $\hat y$ can be assigned as follows:
\begin{equation}
    \hat y = \text{top-C}(p_j^r).
\end{equation}

\subsection{Confidence Penalty Optimization}
We deploy a confidence penalty optimization method to enhance the ability of LoR to capture important local information by penalizing confident output distributions. Specifically, we first obtain the original image representation by passing the image patch embedding through an image encoder. 
Then, concatenated with soft prompts initialized by four description representations to derive the original image representation $H^{o}$, which is fed into the classifier, yielding the original confidence distribution.
We compare the probability changes in reconstructed confidence and original confidence for the correct tag. We calculate the sum of the magnitudes of the decreases in reconstructed confidence relative to original confidence and incorporate it as a penalty into the loss.
\begin{equation}
  \mathcal{L}_{pen}=-\frac{1}{n}\sum^n_{i=1} \sum^m_{j=1}1\{y_i=j\} (e_r^{\theta_j^\top H_i^{r}}-e_o^{\theta_j^\top H_i^{o}}).
\end{equation}
where $e_o^{\theta_j^\top H_i^o}$ is the original confidence distribution. Ultimately, the final loss that needs to be minimized is demonstrated as $\mathcal{L}=\mathcal{L}_{main} + \mathcal{L}_{pen}$.


\begin{table*}[!t]
\centering
\resizebox{\linewidth}{!}{
\begin{tabular}{ccccccccccccc}
\hline  
\multicolumn{1}{c}{\multirow{2}{*}{\textbf{Methods}}} &   \multicolumn{4}{c}{Top-1}  & \multicolumn{4}{c}{Top-3}   & \multicolumn{4}{c}{Top-5} 
\\ \cline{2-13}
\multicolumn{1}{c}{} &  CR &   CF1    &   OR &   OF1&   CR  &   CF1  &   OR &   OF1&   CR  &   CF1  &   OR &   OF1
\\  \hline
ResNet-101  &39.13 &39.20  & 3.21 &1.44&40.19&38.97&3.92&1.08 &47.39 &44.44 & 3.18 &1.27 \\
 CNN  &47.81& 47.33 &1.02 &0.35&48.23&46.45&2.08&0.33& 47.98 &47.77 &0.97& 1.20 \\
ADD-GCN   &49.45& 49.61 & 18.75 &15.19&49.98&50.32&15.83& 13.85& 49.89& 49.89  &13.28& 7.59 \\
ASL  & 49.43& 49.57  &29.02 &16.57&49.67&49.72&21.87& 15.49 &50.12 &50.04& 10.57& 14.09 \\
\hline
Q2L & 31.80& 31.48&8.19 &1.59&32.04&30.89&8.21& 1.47&32.91& 32.51 &8.04& 1.30 \\
TSFormer   &48.63 &46.11 &9.24& 1.91&48.92&46.02&9.10&1.87 &48.86&46.25& 8.25& 1.75 
\\
C-Trans   &  49.89  &  49.83&  9.23  &  2.13 & 49.99 &49.78&8.76&2.11&49.92& 49.56 & 7.43 & 2.05
\\
StickerCLIP  & 49.59  & 50.51  & 1.33 &  1.93  & 50.81&51.05 &3.57  & 3.334  & 51.31  &51.26  &5.63&3.87 \\
CSRA  & 61.42& 60.51 &30.82&23.30&63.23&61.52&31.71&22.25  &64.49 &62.09 &33.57& 22.93 \\
Swin  & 62.74& 63.67& 29.04 & 43.01&65.11&57.83&31.77&31.94&65.42 &55.32  &33.62 &21.39 \\
 BeiT  &62.40&64.68 & 29.91& 43.92&65.42&58.79&32.02& 31.55 &66.11 &55.80  &34.70 &20.69 \\
 CaiT & 62.49 &64.02 & 29.39&43.83&65.66&60.30&31.56& 30.04& 66.18& 56.89 & 33.67 &19.68 \\
DeiT   &62.19 &64.58 & 30.03& 44.75&65.04&61.67&32.31&32.56 &66.15&61.64 &34.38 &24.67 \\
ViT  & 62.20 & 64.58 &30.01 &  44.74 &65.89&62.69&33.03& 33.56&66.85  &62.35  &34.41 &26.34 \\
LORA	&63.45	&65.87&	29.45&	42.97	&66.46	&62.38	&32.42	&33.32	&65.89	&62.75&	34.12&24.37\\
 \hline
MiniGPT4& 8.97 & 14.89 &  1.42& 2.31 & 8.80 & 14.32 &2.78  & 2.07 &8.69&13.23&2.99&1.74  \\
LLaVA& 9.22 & 15.32 & 1.56 &2.45  & 9.45 & 15.43 & 3.08 &2.54&9.55&14.07& 3.02&2.74 \\
Qwen-VL& 9.43 & 15.81 &  1.76&2.58 & 9.98 &  16.19& 3.04 & 3.24 &10.25&16.07&3.65&3.08  \\ \hdashline
\rowcolor{gray!20} \textbf{Ours}  & \textbf{68.69$^\ast$} & \textbf{70.14$^\ast$}   &  \textbf{32.50$^\ast$} & \textbf{48.21$^\ast$} &\textbf{70.17$^\ast$}&\textbf{68.39$^\ast$}&\textbf{37.19$^\ast$}&\textbf{36.60$^\ast$} &  \textbf{71.85$^\ast$} &\textbf{66.70$^\ast$} & \textbf{38.72$^\ast$} & \textbf{28.87$^\ast$}\\
\multicolumn{1}{r}{-w/o LoR} & 64.32 & 67.04 & 30.43&  45.29&66.13&66.69&34.16&34.10 & 68.01&  63.39&  35.86&  25.86  \\
\multicolumn{1}{r}{-w/o prompt} &65.01& 68.91 & 31.05& 46.36&67.98&67.02&35.22& 34.96&69.73& 64.62& 36.99& 27.56  \\
\multicolumn{1}{r}{-w/o penalty} & 67.02& 68.32 & 31.91& 46.96&68.12&67.89&36.30&35.67& 69.64 &64.71 & 36.67& 27.38   \\
\hline
\end{tabular}
}
\caption{Performance (\%) of various methods in our StickerTAG.  \textbf{Bold} indicates the model with the best performance. We assert significance $^\ast$ if $p$-value < 0.05 under a t-test with the baseline models. w/o means without.}
\label{tab2}
\vspace{-0.2cm}
\end{table*}

\section{Experiment}
\subsection{Experimental Set-up}
\textbf{Implement details.}  
For data participation, we divide our StickerTAG dataset into a training set, a validation set, and a testing set with a ratio of 8:1:1.
Att$^2$PL employs Swin Transformer \cite{24-vaswani2017attention} as the image encoder, we apply a $1 \times 1$ convolution layer resulting in an output dimension of $3072 = 32 \times 32 \times 3$.
For attribute-oriented description generation, we employ a 12-layer Multilingual BERT \cite{23-kenton2019bert} 
as the text encoder.
We utilize Qwen-VL \citep{44-bai2023qwen} for attribute-oriented description generation.
For a fair comparison with other methods, we resized all images to $H \times W = 224 \times 224$ as input resolution in both training and test phases throughout all experiments. 
The optimization of Att$^2$PL is done by AdamW \cite{28-kingma2014adam} with a learning rate of $1e^{-5}$, weight decay $1e^{-2}$, and the batch size 8 for maximally 20 epochs. All experiments are performed on Nvidia RTX-3090Ti GPU and our model is implemented in PyTorch \footnote{http://xxx.com}.

\textbf{Metrics.} 
To comprehensively evaluate performance, we adhere to the methodology of prior studies \cite{45-wang2016cnn,46-ge2018multi, 47-chen2019multi} and present metrics including average per-class recall (CR), F1 (CF1), and the average overall recall (OR), F1 (OF1). Additionally, we include results for top-1, top-3, and top-5 tags. Higher values across all metrics indicate superior performance. Tags for each sticker image are deemed positive if their predicted probabilities surpass 0.5.

\subsection{Compared Methods}
We adopt comprehensive mainstream models for multi-tag recognition, building upon our proposed StickerTAG dataset: 
(1) \textit{Conventional methods}: CNN \cite{29-krizhevsky2012imagenet}, ResNet-101 \citep{30-he2016deep}, ADD-GCN 
 \citep{31-ye2020attention}, ASL \citep{32-ridnik2021asymmetric}. 
(2) \textit{Transformer-based methods}: Swin \cite{33-liu2021swin}, ViT \cite{34-osovitskiy2020image}, BeiT \cite{35-bao2021beit}, DeiT \cite{36-touvron2021training}, CaiT \cite{37-touvron2021going}, CSRA \cite{38-zhu2021residual}, Q2L \cite{39-liu2021query2label}, TSFormer \cite{40-zhu2022two}, C-Tran \cite{41-lanchantin2021general}, StickerCLIP \cite{16-zhao2023sticker820k}, and LORA \cite{15-liu2022ser30k}. 
(3) \textit{Multi-modal large language methods}: Mini-GPT4 \cite{42-zhu2023minigpt}, LLaVA \cite{43-liu2023visual}, Qwen-VL \cite{44-bai2023qwen}.
For MLLMs, we list the tag set, present the sticker image, and prompt them with " Please select the most suitable tags for the above sticker."

\subsection{Main Results}
We examine the performance of our model and baselines in terms of each evaluation metric, as shown in Table \ref{tab2}. 
The experimental results indicate that the Att$^2$PL approach largely achieves the best results. Moreover, the significance tests of our Att$^2$PL over the baseline models demonstrate the effectiveness of our proposed method, presenting a statistically significant improvement based on most evaluation metrics with $p$-value < 0.05. We also observe that the performance of Top-1 is better than Top-5, and overall it is inferior to individual categories. This indicates that in a multi-tag scenario, the model finds it challenging to accurately select all tags correctly, highlighting the inherent difficulty of this task. 

\textbf{Analysis of Transformer-based methods.}  Transformer models are based on the Transformer architecture, which is known for its self-attention mechanism, allowing the model to focus on different parts of the input sequence and weigh their importance concerning the current context. 
CSRA, the strongest baseline, generates class-specific features using a spatial attention score combined with class-agnostic average pooling, effectively capturing spatial distributions. 
However, these approaches are not well-suited for the multi-tag sticker recognition task due to their lack of consideration for sticker-specific characteristics, such as fine-grained visual and semantic features. 

\textbf{Analysis of multi-modal large language methods.}
We find that multi-modal large language models (MLLMs) perform suboptimally, particularly with long texts and numerous tags. MLLMs often struggle to fully grasp context and tend to prioritize initial tags, overlooking subsequent ones. This results in poor performance in multi-label tasks. 
In the future, exploring ways to leverage MLLMs to enhance the performance of multi-tag or extreme multi-label recognition tasks is a worthwhile pursuit. Strategies such as employing larger models for data augmentation or incorporating reasoning explanations could be considered.

\textbf{Experiments on SER30K Datasets.}
We also apply our Att$^2$PL method on the existing SER30K dataset, as shown in Table \ref{tab_ser30k}. LORA \cite{15-liu2022ser30k} captures local information in stickers through the attention mechanism. In contrast, our method not only focuses on local information but also explores the characteristics of stickers from different perspectives, delving deeper into their features to enhance classification performance.

\subsection{Ablation Study}
To understand the influence of each component of our framework, we further conduct an ablation study.
As shown in Table \ref{tab2}, the performances of all ablation models are worse than that of Att$^2$PL under all metrics, which demonstrates the necessity of each component in Att$^2$PL.
The most significant decrease in performance occurs when excluding LoR, highlighting its paramount importance in our method. LoR plays a crucial role in guiding the model to focus more on the crucial information within stickers. Despite the performance drop without LoR, the model still outperforms other baselines, indicating that both prompt and penalty contribute to enhancing model performance to a certain extent. Furthermore, the thoughtful design of the prompt, incorporating sticker-specific features, contributes to the robustness of our model. This suggests that attributes of stickers play a pivotal role in the effective tag recognition of stickers.

\begin{table}[!t]
\centering
\begin{tabular}{cc}
\hline  
\textbf{Methods}  &  \textbf{Accuracy}\\ \hline
TFN	\cite{49-zadeh2017tensor} &54.19\\
MCB \cite{50-gao2016compact}&	58.18\\
PDANet-T \cite{51-zhao2019pdanet}&	68.93\\
WSCNet-T \cite{52-yang2018weakly}&69.45\\
LORA \cite{15-liu2022ser30k}&\textbf{}{70.73}\\ \hdashline
\textbf{Ours} &\textbf{72.86}\\
\hline
\end{tabular}
\caption{Performance (\%) of the recently proposed emotion recognition methods and our method on the SER30K dataset. }
\label{tab_ser30k}
\vspace{-0.2cm}
\end{table}

\subsection{Effect of Local Re-attention Module}
We further analyzed the feasibility and effectiveness of the LoR module from two perspectives and presented the results in Table \ref{tab3}. Initially, we replaced the LoR module with commonly used attention mechanisms, including spatial attention. We observed a slight degradation in model performance. Notably, the self-attention mechanism exhibited the poorest performance. This is attributed to the fact that our LoR design is based on the Swin Transformer, which inherently performs self-attention calculations on image patches. In comparison, spatial attention and channel attention showed relatively better performance.

We also explored various image encoders in LoR. It can be observed that our model achieves better performance when using ViT. This might be due to the ViT's ability to capture global contextual information through self-attention in each image patch's embedding, enhancing the model's overall global perception when processing the entire image. Additionally, the LoR module designed in this paper is configured to deepen its focus on specific features and regions related to stickers, contributing to improved results.
Furthermore, despite CaiT showing the least favorable performance as the image encoder, when considering this observation alongside Table \ref{tab2}, it becomes evident that our approach outperforms other baseline models.

\begin{table*}[!t]
\centering
\resizebox{0.9\linewidth}{!}{
\begin{tabular}{ccccccccccccc}
\hline  
\multicolumn{1}{c}{\multirow{2}{*}{\textbf{Methods}}} &   \multicolumn{4}{c}{Top-1}  & \multicolumn{4}{c}{Top-3}   & \multicolumn{4}{c}{Top-5} 
\\ \cline{2-13}
\multicolumn{1}{c}{} &  CR &   CF1    &   OR &   OF1&   CR  &   CF1  &   OR &   OF1&   CR  &   CF1  &   OR &   OF1
\\  \hline

\textbf{Ours}   &  {71.85} &\textbf{66.70} & {38.72} & {28.87}&{68.69}&	{70.14}	&{32.50}&	{48.21}	&{70.17}	&{68.39}	&{37.19}	&{36.60}\\ \hdashline
\rowcolor{gray!20}  \multicolumn{13}{l}{\textit{Different attention mechanism compared with LoR}}  \\
\multicolumn{1}{c}{Spatial} & 69.44&  62.88 &  37.87 & 26.51  &	66.98&	68.45	&31.20	&47.01&	68.13&66.43&	35.38	&34.78  \\
\multicolumn{1}{c}{Channel}  & 68.66& 62.15& 35.33 &24.76&65.64&	67.58	&30.64&	46.13	&67.17	&65.84	&35.03&	33.64     \\
\multicolumn{1}{c}{Self} &  68.30& 62.08& 35.65 &24.92  &65.39&	66.76	&30.13	&45.38	&67.52	&65.39&	34.95&	33.51  \\ \hdashline
\rowcolor{gray!20}  \multicolumn{13}{l}{\textit{Different backbone in LoR}}\\
CaiT&69.97& 63.31 & 39.04&27.40&	67.03&	68.64	&31.25	&47.13	&68.34	&66.59&	35.99	&35.01\\
BeiT&71.32 &65.54 & 43.23 &31.56 &67.81&	69.42&	32.08&	47.76&	69.79&	67.85	&37.08&	35.97\\
DeiT&72.81 &65.60 & 44.38& 32.04 &68.99	&70.07	&32.34&	48.25	&69.94&	68.54&37.03&36.64\\
ViT& \textbf{73.31} &65.81 &\textbf{45.08} &\textbf{32.42} &	\textbf{69.32}&\textbf{70.54}	&\textbf{32.58}&	\textbf{48.34}	&\textbf{70.21}	&\textbf{68.75}&	\textbf{37.47}&	\textbf{37.02}
\\
\hline
\end{tabular}
}
\caption{Effect of Local Re-attention Module.  
"Spatial", "Channel", and "Self" represent spatial attention, channel attention, and self-attention, respectively.}
\label{tab3}
\end{table*}

\begin{figure*}[!t]
  \centering
  \includegraphics[width=0.9\linewidth]{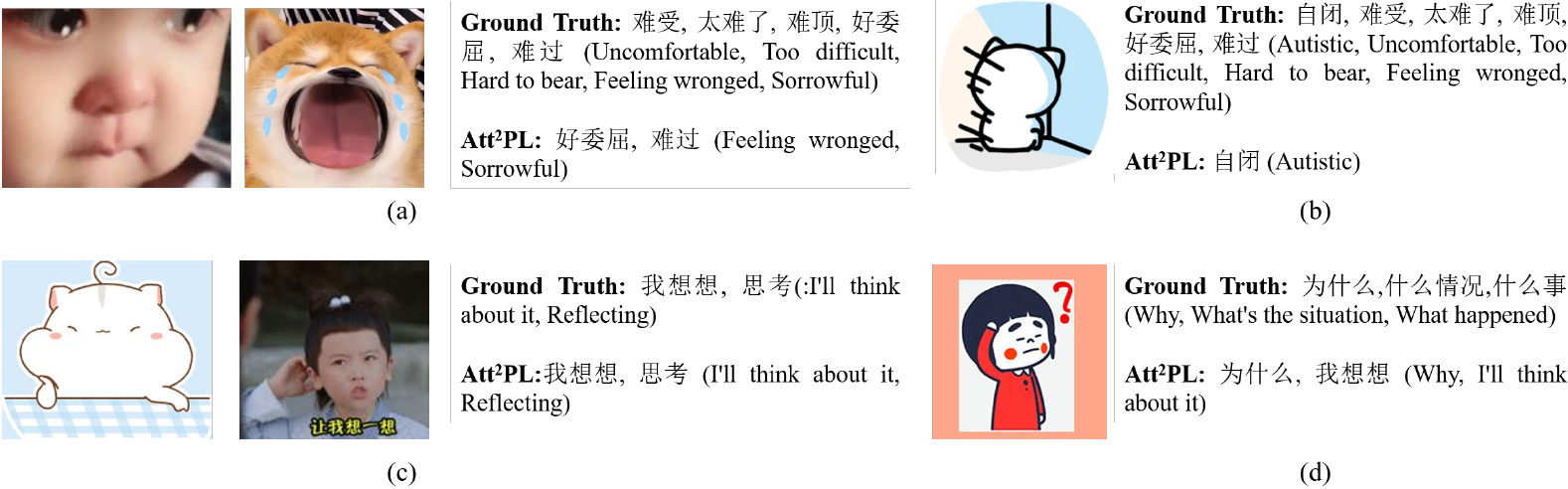}
  \caption{Examples of stickers with ground truth tags and the predicted tags inferred by our Att$^2$PL framework.}
  \label{f5}
  \vspace{-0.2cm}
\end{figure*}

\subsection{Case Study}
Several cases with ground truth tags and predicted tags inferred by our Att$^2$PL approach are depicted in Figure \ref{f5}.
Upon observation of examples (a) and (c), it becomes evident that sticker features are predominantly manifested in actions and expressions, establishing a direct correlation with the associated tags. For instance, a crying expression could be associated with tags like \textit{"I'm too distressed"} or \textit{"Feeling wronged"}, while the act of holding one's head and a contemplative facial expression may indicate tags such as \textit{"Thinking"} or \textit{"I'll think about it"}. The proposed method efficiently guides the model's attention toward these crucial features, thereby improving recognition performance.

\textbf{Error analysis.} We also observed instances of incomplete recognition in example (b) of Figure \ref{f5} with our Att$^2$PL model. 
While it successfully infers the sticker as the \textit{"Autistic"} tag, it struggles to recognize other tags in more depth.
This challenge primarily arises from learning the similarities between different tags in multi-tag sticker recognition. 
Furthermore, for similar actions like the one representing the question \textit{"Why"} in example (d), the model may mistakenly recognize it as the tag \textit{"I'll think about it"}. This highlights the difficulty in distinguishing between similar tags and expressions, which is inherent in this task.

\section{Conclusion}
Stickers can be interpreted differently by users in the real-world scenario, leading to a variety of associated tags. 
This paper introduces a StickerTAG dataset, comprising 461 tags and 13,571 sticker-tag pairs. Additionally, we present the Att$^2$PL framework, specifically designed to tackle the distinct challenges of multi-tag sticker recognition. 
We first obtain the attribute-oriented descriptions from stickers based on four attributes. Next, we introduce a local re-attention module to focus on local information, followed by the application of prompt learning to guide the recognition process.
Furthermore, we utilize confidence penalty optimization to penalize the confident output distributions. 
Our Att$^2$PL method outperforms existing methods on our StickerTAG and SER30K dataset, emphasizing the necessity of incorporating stickers' distinctive characteristics into recognition tasks.

\section*{Limitations}
Although the proposed approach yields promising results, there are still some limitations to be addressed, including:
1) Stickers exhibit diverse styles and expressions in real-world conversations, which may impact the effectiveness of tag recognition.
2) Distinguishing between similar tags remains a challenge, particularly in cases where subtle nuances in meaning are present.
There is room for improvement in the model's fine-grained recognition capabilities, particularly in distinguishing between closely related concepts.
Future research could delve into the stylistic variations of stickers and explore additional strategies to address the complexities of tag recognition, thereby enhancing the overall performance of the system.

\section*{Ethics Statement}
The original copyright of all stickers belongs to the respective owners, and they are publicly available for academic use. The sticker sets can be freely accessed online. The annotations' copyright belongs to our group, and they will be released to the public free of charge. After consulting with legal advisors, the StickerTAG dataset is freely accessible online for academic purposes. However, commercial use and distribution to others without permission are strictly prohibited.

Our data collection process includes manual annotation. The annotated conversation corpus and sticker sets do not contain any personally sensitive information. The annotators received fair compensation for their annotation work.

\bibliography{acl_latex}

\end{document}